# Deep CNN for Coherent Seismic Noise Removal: A Perspective

**Authors**

Rohit Shrivastava, Ashish Asgekar (Shell Technology Center, Bengaluru, India)

Evert Kramer (Shell Technology Center, Amsterdam)

| | |
|---|---|
| Title | Deep CNN for Coherent Seismic Noise Removal: A Perspective |
| Paper Number | 2068 |
| Paper Status | Accepted for oral |
| Session Details | ML in Seismic Processing 1 - Noise<br>Wednesday, Jun 10, 2020<br>1:30 PM - 5:10 PM<br>Elicium 1 |
| Presenting Author | Rohit Shrivastava<br>Affiliations: Shell Technology Center |
| Co-Author | Dr Ashish Asgekar<br>Affiliations: Shell Technology Center |
| Co-Author | Evert Kramer<br>Affiliations: Shell Technology Center |

**Introduction**

Seismic data contain different kinds of noise, their source may be artificial, such as noise of acquisition ships, or natural, such as swell noise or receiver noise (Bormann and Wielandt, 2002); they can be classified into two different categories: coherent and random. Denoising seismic data or "seismic denoising" is an important processing step, which consumes a significant amount of time, whether it is for Quality control or for the associated computations (esp. coherent noise attenuation, Khalil et al. (2013)).

Typically, most noise attenuation methods rely on the assumption of Gaussian-distribution of the noise and the predictability of coherent signals. These methods transform data into a convenient domain where the signal can be represented by sparse set of features, well separated from noise, such as $f - x$ or $\tau - p$ domains. One then needs to determine an optimum filter/nonlinear mapping function to remove coherent noise pattern from trace samples (Peng, 2014; Zhu et al., 2019). The noise attenuation mechanisms employed for Surface Related Multiple Extraction (SRME) or removal of Seismic Interference (SI) are often based on these principles (Verschuur and Berkhout, 1993; Jansen et al., 2013). The parameters of the aforementioned nonlinear mapping function are manually set by the processor using trial and error or past experience.

Finding optimal parameters for filtering is far from trivial, with several limitations due to limited wavefield sampling, frequency content of the measured signal and unknown source wavelet. Over and above this, there is a human component which is prone to error in finding optimum filter parameters and also adds subjectivity in processing. Despite decades of development, such deterministic methods were found to be limited in noise suppression.

Machine learning is at its essence an optimization scheme where the loss function between desired output and forward modeled output is minimized. Neurons in multiple layers of a neural network, such as a Convolutional Neural Network (CNN) "learn" the functional form of filters between the input and output images (LeCun et al., 1998). In this manuscript, we present results of our work in training CNN for denoising seismic data, specifically attenuation of surface related multiples and removal of overlap of shot energies during simultaneous-shooting survey. The proposed methodology is being explored not only for its ability to minimize human involvement but also because of the trained filter's ability to accelerate the process, hence, reduce processing time. In the section we discuss details of our methods, followed by results and discussion.

**Theory: Denoising as a Deep Learning Problem**

Denoising images using signal processing is a well known subject of study and is at the core of machine vision (Buades et al., 2005). Many image denoising technologies have been developed in recent years, but the problem to maintain the structural integrity of the image still remains along with the non-convex nature of these methods (require manual tuning of parameters Zhu et al. (2019)). Supervised learning techniques like neural networks have shown a lot of promise in denoising images without prior knowledge of noise level and the type of noise as long as training data contain both noisy and reference images (Lemarchand et al., 2019; Bekara, 2019). Multiple layers of network architecture is the reason behind the terminology "deep". Many network architectures were employed in the past decade that have reduced computational time significantly for certain quality of image processing of certain standard datasets (LeNet, ResNet, AlexNet, RNN, etc.). Using deep learning for seismic denoising has multi-fold advantages over conventional denoising techniques– CNN methodology as mentioned previously does not require prior knowledge of noise level or the type; once trained the computational expense of the filtering action of the network is very low. Further, denoising operation can be carried out in the shot domain itself making a domain transformation redundant, hence, enabling real time denoising (Slang et al., 2019). The same network can be trained in such a manner that it can perform multiple denoising operations simultaneously, thus reducing the processing time even further.

**Method: CNN-Unet**

Our network architecture is based CNN reported in the literature Ronneberger et al. (2015) and Quan et al. (2016). It consists of an encoding path to retrieve features of interest and a symmetric decoding path that recreates predicted output. Encoding and decoding parts contain multiple hidden layers of neurons over multiple levels of smoothing. These will extract different features that occur over various spatio-temporal scales. The layers are constructed using convolution layer, a rectilinear unit and batch normalization in succession. Convolution layers are followed by max-pooling layers located between different levels in the encoding path; they reduce image dimensions and result in feature compression. In the decoding path, we use deconvolution layers to up-sample image data using interpolation. One half of an upsampled image is concatenated with feature matrix from the layer in the encoding path at the same level and the result is convolved further to create next level in the decoding path. "Skip connections" employed in this network architecture (unlike a fully connected CNN) assists in improving the convergence of training and prediction performance Zhu et al. (2019).

To train the CNN for SRME we used synthetic traces generated with different velocity models, a few with 2 reflecting layers and other models constructed from various parts of the Marmousi model. For each subsurface model we synthesize traces containing primaries and multiples ($P+M$) using reflecting boundary conditions, and primaries-only ($P-only$) using absorbing boundary conditions. We convolved traces in the latter data for source and receiver ghosting. We also obtain estimates of multiples using SRME (Verschuur and Berkhout (1993); $M\prime$). CNN input layer contains two channels: one with single-shot panel constructed using reflecting boundary condition: "$P+M$", while the other channel contains SRME estimate of multiples for the same shot panel: "$M\prime$". The output channel contains the same shot-domain panel of traces synthesized using an absorbing boundary condition.

For SI removal/deblending seismic data, the receptive field for the network is kept as large as possible, "patch-size" maximum, to avoid loss of structural integrity of features (same was done for SRME as well.) CNN is trained to remove seismic interference occurring on the right side of the image, left part of the repeated pattern is retained as signal. Training data were synthesized using flat 2-layer model and marmousi model as discussed below. Details of U-Net architecture and training methodology with "patching" and "batching" of data will be a subject of separate communication under preparation.

**Results: Seismic Interference Removal**

Figures 1 and 2 represent seismic responses of two-layer subsurface model and Marmousi model respectively when two sources are excited simultaneously with and without time dithering (left source is considered as signal and the right source is considered as noise). Fig. 1(c), 1(d), 2(c) are the deblended seismic traces obtained by using deep CNN filters, these images are not a part of the training datasets. Deep CNN filter trained with low epoch number leaves residual noise (right source's response) in the filtered image, which are random in nature.

**Results: SRME**

We employed CNN U-net for suppression of multiples in synthetic data and our network converged after about 800 epochs of training using about 2000 images. Afterwards we tested the trained network using an independent subset of input data. Figure shows the result of deep network used to suppress multiples. Typically network error reduces by a factor of 500-1000 after 1000 epochs of training and even fainter desired features are largely preserved while removing bright features nearby.

**Conclusions**

Neural network methodology for seismic denoising is a robust operation. It can carry rapidly out multiple denoising processes while being computationally efficient, thus significantly reducing the time for seismic processing. Network works well without prior knowledge of noise level and even when signal and noise occupy the same frequency bandwidth. The Unet architecture of CNN with its skip connection property accelerates the convergence during training process for both problems.

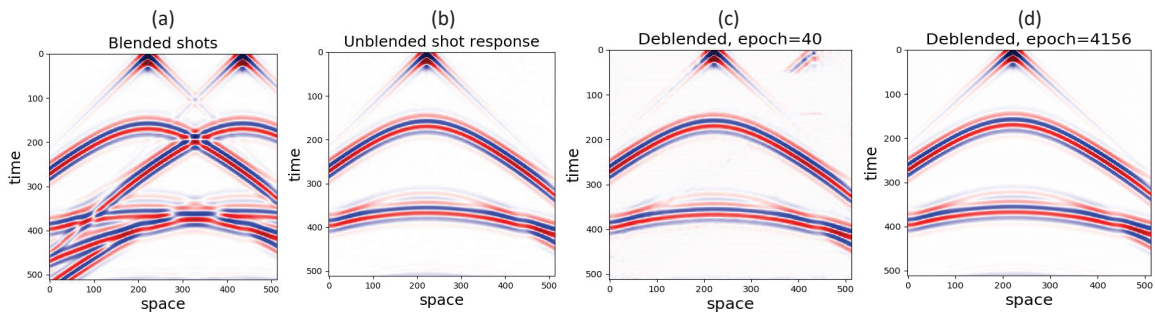

*Figure 1* Seismic response of a two layer subsurface model (with water on the top) when two sources are excited simultaneously with Ricker wavelet excitation source (central frequency 35 Hz): (a)Blended response of simultaneous shots (b)Unblended response when the second source is not excited (c) De-blended seismic response obtained by using deep CNN (neural network trained with less number of epochs, hence higher value of mean square logarithmic error loss function) (d) Deblended seismic response with higher number of epochs.

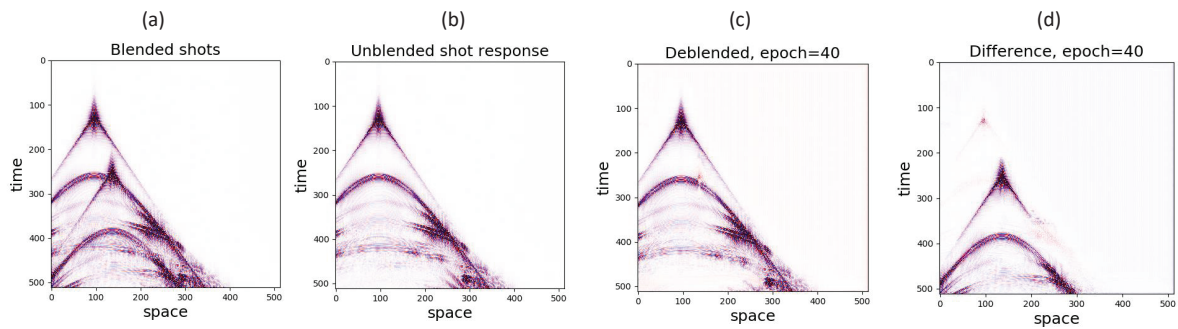

*Figure 2* Seismic response of Marmousi model with water layer on the top and time dithering.

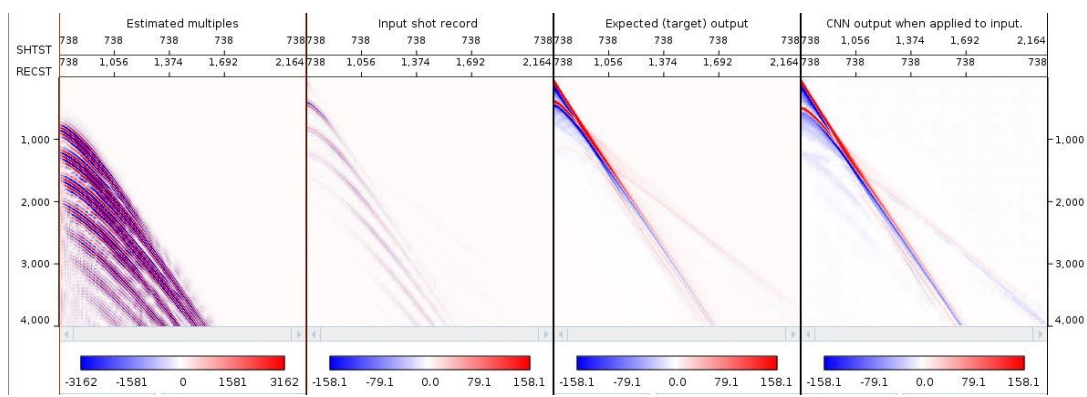

*Figure 3* CNN applied to SRME using synthesized traces: Panels from left to right show predicted multiples of one input shot panel (channel 2) generated with reflecting boundary condition, the input shot panel itself (channel 1), the corresponding shot record synthesized using absorbing boundary condition (target) and the output of the CNN-Unet when applied to two input channels as described in the text. Vertical scale is in milliseconds and colorbars for values are displayed at the bottom.

For SI removal, low epoch numbers result in poor attenuation, whereas a network trained with high number of epochs gives a higher SNR image after removal. Intelligent choice of training data sets can result in faster convergence rate and better SNR of image after SI removal for low epoch numbers. However, the residual parts of the noise after SI removal (Fig. 2(c)) by the network can be stacked out due to its random nature.

Ubuquity of the trained filter can be increased by incorporating more seismic responses of different models in the training datasets. Future work will include studying robustness of networks to artefacts in the output, generalization of networks to deploy them on trace data from new subsurface models and deployment of a different network architectures to exploit the incoherent nature of noise in channel domain for better convergence during network training.

## Acknowledgements

We would like to acknowledge Akshat Abhishek, Fons ten Kroode, Gautam Kumar, Koos de Vos, Wim Walk and Chris Willacy for support and discussions.